\documentstyle[amstex,amssymb,12pt,righttag]{article}

\newenvironment{Proof}{\noindent {\it Proof:\ }}%
{\hfill {\mbox{$\Box$}} \par  \vspace{1.5ex}}

\newenvironment{Rem}{\noindent {\it Remark.\ }}%
{\hfill \par \vspace{1.5ex}}

\newtheorem{Th}{Theorem}%[section]  
\newtheorem{Prop}{Proposition}%[section]
\newtheorem{Lem}{Lemma}%[section]
\newtheorem{Cor}{Corollary}%[section]
\newtheorem{Def}{Definition}%[section]

\newcommand{\EE}{{\Bbb E}}

\newcommand{\RR}{{\Bbb R}}
\newcommand{\TT}{{\Bbb T}}
\newcommand{\SS}{{\Bbb S}}
\newcommand{\VV}{{\Bbb V}}

\newcommand{\ZZ}{{\Bbb Z}}

\newcommand{\cF}{{\cal F}}
\newcommand{\cI}{{\cal I}}

\newcommand{\cR}{{\cal R}}
\newcommand{\cQ}{{\cal Q}}

\newcommand{\calC}{{\cal C}}

\newcommand{\St}{\operatorname{St}}                          
\newcommand{\opP}{\operatorname{P}}
\newcommand{\const}{\operatorname{const}}                          

\newcommand{\ba}{{\boldsymbol a}}
\newcommand{\bx}{{\boldsymbol x}}
\newcommand{\bN}{{\boldsymbol N}}

\newcommand{\br}{{\boldsymbol r}}
\newcommand{\by}{{\boldsymbol y}}

\newcommand{\bn}{{\boldsymbol n}}

\newcommand{\bX}{{\boldsymbol X}}

\newcommand{\bY}{{\boldsymbol Y}}

\newcommand{\bphi}{{\boldsymbol \phi}}
\newcommand{\bPhi}{{\boldsymbol \Phi}}
\newcommand{\bpsi}{{\boldsymbol \psi}}
\newcommand{\bPsi}{{\boldsymbol \Psi}}

\newcommand{\gl}{{\frak l}}

\newcommand{\tB}{\tilde{B}}
\newcommand{\tC}{\tilde{C}}

\newcommand{\tbx}{\tilde{\boldsymbol x}}
\newcommand{\tby}{\tilde{\boldsymbol y}}

\newcommand{\ccR}{{\cal R}^\circ}

\newcommand{\vbx}{\vec{\boldsymbol x}}
\newcommand{\D}{{\Delta}}

\newcommand{\ra}{\rightarrow} 

\begin{document}

\title{\sc Quadratic reductions of quadrilateral lattices} 

\author{Adam Doliwa \\ \\
Istituto Nazionale di Fisica Nucleare, Sezione di Roma\\
P-le Aldo Moro 2, I--00185 Roma, Italy \\ \\
Instytut Fizyki Teoretycznej, Uniwersytet Warszawski\thanks{permanent 
address}\\
ul. Ho\.{z}a 69, 00-681 Warszawa, Poland}

\date{}

\maketitle

%E-mail: {\tt doliwa@roma1.infn.it, doliwa@fuw.edu.pl}

\begin{abstract}

\noindent It is shown that 
quadratic constraints are compatible with the geometric integrability scheme
of the multidimensional quadrilateral lattice equation. The corresponding
Ribaucour-type
reduction of the fundamental transformation of quadrilateral lattices 
is found as well, and superposition of the Ribaucour transformations 
is presented in the vectorial framework.
Finally, the quadratic reduction approach is illustrated on the
example of multidimensional circular lattices. 

\medskip

{\it PACS:} 0340K, 0240

\end{abstract}

\newpage

\section{Introduction}
The connection between differential geometry and modern theory of
integrable partial differential equations has been observed 
many times~\cite{Sym,Dryuma,Tsarev,Bobenko2b2}. Actually, a lot of basic 
integrable systems, like the sine-Gordon, Liouville, Lam\'e or Darboux
equations, were studied by distinguished geometers of the 
XIX-th century~\cite{Lame,Bianchi,DarbouxOS} (see 
also~\cite{Eisenhart-TS,Lane}). 

It turns out that
many integrable systems of a geometric origin are
reductions of the Darboux equations, which describe submanifolds in 
$\RR^M$ parametrized by conjugate coordinate systems 
(conjugate nets)~\cite{DarbouxOS}.
The Darboux equations were rediscovered and solved, in the matrix 
generalization,
in~\cite{ZakMa} using the $\bar\partial$--dressing method
(for another approach to the Darboux equations see~\cite{GrZel1,GrZel2}
and references therein). 
More recently it was shown in~\cite{DMMMS} that classical 
transformations of the 
conjugate nets, which are known as the Laplace, L\'evy, Combescure,
radial and fundamental transformations~\cite{Eisenhart-TS,Lane}, 
provide an interesting geometric interpretation of the basic
operations associated with the multicomponent Kadomtsev--Petviashvilii 
hierarchy. 

The distinguished reduction of the Darboux equations is given by the
Lam\'e equations~\cite{Lame,DarbouxOS,Bianchi}, which describe 
orthogonal systems of coordinates; these equations were
solved recently in~\cite{Zakharov}. The reduction of the fundamental
transformation compatible with the orthogonality constraint is 
provided by the Ribaucour transformation~\cite{Bianchi},
which vectorial generalization was constructed in~\cite{LiuManas}.

During last few years the connection between geometry and integrability
was observed also at a discrete level~\cite{BP1,DS-AL,BP2,BS}. 
In particular, in~\cite{MQL}
it was shown that the integrable discretisation of the conjugate
nets is provided by multidimensional quadrilateral lattices (MQL), 
i.e., maps $\bx: \ZZ^N \ra \RR^M$ with all the elementary quadrilaterals 
planar (see also~\cite{Sauer,DCN}). 
The geometrically distinguished reduction
of quadrilateral lattices are multidimensional circular lattices (MCL),
for which all the elementary quadrilaterals should be inscribed in 
circles~\cite{Bobenko-DO,CDS}. The circular lattices provide
the integrable discretization of the orthogonal coordinate systems of 
Lam\'e. 
In \cite{CDS,DMS} it was demonstrated that the circularity constraint is 
compatible 
with the geometric and analytic
integrability scheme of MQL and provides an integrable reduction 
of the corresponding equations. 

Also the Darboux-type transformations~\cite{ms} of quadrilateral lattices have
been studied from various points of view \cite{DCN,MDS,KoSchief2,TQL}. 
In~\cite{TQL} the general theory of transformations applicable
to any quadrilateral lattice was presented, and all the classical
transformations of conjugate nets have been generalized to a discrete
level. In~\cite{KoSchief2} there was found, among others, 
the (discrete analog of the) Ribaucour 
transformation compatible with the circularity constraint.

In conclusion of \cite{CDS} there were given some arguments supporting 
the conjecture that a more general then circularity, but still
{\it quadratic}, constraint imposed on the quadrilateral lattice
preserves their integrability scheme. 
In the present paper we develop this observation and prove the general
theorem about the integrability of quadratic reductions of the
MQL equation (Section \ref{sec:quadratic}). 
The corresponding Ribaucour reduction of the fundamental (binary Darboux)
transformation is found in Section \ref{sec:Ribaucour}, where we present also 
superposition of the Ribaucour transformations in the vectorial framework. 
Section~\ref{sec:R-MCL} provides
an exposition of the multidimensional circular
lattices and their Ribaucour transformation from the quadratic 
reduction point of view.

\section{Integrability of quadratic reductions} \label{sec:quadratic}
We recall that planarity of elementary quadrilaterals of 
the lattice $\bx: \ZZ^N \ra \RR^M$ can be expressed in terms
of the Laplace equations~\cite{DCN,MQL}
\begin{equation}  \label{eq:Laplace}
\D_i\D_j\bx=(T_i A_{ij})\D_i\bx+
(T_j A_{ji})\D_j\bx \; \; ,\;\; i\not= j, \; \; \;  i,j=1 ,\dots, N \;  ,
\end{equation}                                                                     
where the coefficients $A_{ij}$, for $N>2$, due to 
the compatibility condition 
of (\ref{eq:Laplace}), satisfy the MQL equation~\cite{BoKo,MQL}
\begin{equation} \label{eq:MQL-A}
\D_k A_{ij} =
(T_jA_{jk})A_{ij} +(T_k A_{kj})A_{ik} - (T_kA_{ij})A_{ik},
\;\; i\neq j\neq k\neq i ;
\end{equation} 
in the above formulas $T_i$ is the shift operator in the $i$-th
direction of the lattice and $\D_i=T_i-1$ is the corresponding
partial difference operator.

In this paper we study lattices $\bx$ contained in a 
quadric hypersurface $\cQ$ of $\RR^M$, $N\leq M-1$.
This additional constraint implies that the lattice points $\bx$ satisfy the 
equation of the quadric $\cQ$, which we write in the form
\begin{equation} \label{eq:quadric}
\bx^t  Q  \bx +  \ba^t \bx + c = 0  \; \; \; ; 
\end{equation}
here $Q$ is a symmetric matrix, $\ba$
is a constant vector, $c$ is a scalar and $^t$ denotes transposition.

Let us recall (see \cite{MQL} for details) that the construction scheme
of a generic $N$-dimensional quadrilateral lattice ($N>2$) 
involves the linear operations only, and is a consequence of the 
planarity of elementary quadrilaterals (or the Laplace equations
(\ref{eq:Laplace})). 
\begin{Th} \label{th:constrMQL}
The point $T_{i}T_{j}T_{k}\bx$ of the lattice is the intersection point 
of the three planes $\VV_{jk}(T_{i}\bx) = 
\langle T_i\bx , T_iT_j\bx , T_iT_k\bx \rangle $, 
$\VV_{ik}(T_j\bx) = \langle T_j\bx , T_iT_j\bx , T_jT_k\bx \rangle $ and 
$\VV_{ij}(T_k\bx) = \langle T_k\bx , T_iT_k\bx , T_jT_k\bx \rangle$ in the 
three 
dimensional space $\VV_{ijk}(\bx)= 
\langle \bx, T_i\bx , T_j\bx , T_k\bx \rangle $.
\end{Th}
\begin{Rem} The above construction scheme  is the geometrical counterpart
of MQL equations (\ref{eq:MQL-A}) and it is called therefore, the {\it
geometric integrability scheme}. It implies, in particular, that the lattice 
$\bx$
is completely determined once a system of 
initial quadrilateral surfaces has been given~\cite{MQL}.
\end{Rem} 
As it was proposed in \cite{CDS}, a geometric constraint in order
to be integrable  must ``propagate" in the construction
of the MQL, when satisfied by the initial surfaces.  
The (geometric) integrability of the quadratic 
reductions is an
immediate consequence of the following classical {\it eight points theorem}
(see, for example \cite{Pedoe}, pp. 420,424).
\begin{Lem} \label{lem:8points}
Given eight distinct points which are the set of intersections of three 
quadric surfaces, all quadrics through any subset of seven of the points must 
pass through the eight point.
\end{Lem}
\begin{Prop} \label{prop:quadr-int}
Quadratic reductions of quadrilateral lattices
are compatible with geometric integrability scheme
of the multidimensional quadrilateral lattice equation.
\end{Prop}
\begin{Proof}
Since the construction of the MQL for arbitrary $N\geq3$  can be 
reduced to the compatible construction of three 
dimensional quadrilateral lattices~\cite{MQL}, it is enough to 
show that the constraint is preserved in a single step 
described in Theorem \ref{th:constrMQL}.
We must show that if the seven points $\bx$, $T_i\bx$, $T_j\bx$, $T_k\bx$,
$T_iT_j\bx$, $T_iT_k\bx$ and $T_jT_k\bx$ belong to the quadric $\cQ$,
then the same holds for the eight point $T_iT_jT_k\bx$ as well. 
Denote by $\cQ_{ijk}(\bx)$ the intersection of the quadric $\cQ$ with the
three dimensional space $\VV_{ijk}(\bx)$, there are two possibilities:\\
i) $\cQ_{ijk}(\bx)=\VV_{ijk}(\bx)$, or \\
ii) $\cQ_{ijk}(\bx)\subset \VV_{ijk}(\bx)$ is a quadric surface.\\
Since in the first case the conclusion is trivial, we concentrate on the 
second point. Recall that two planes in $\VV_{ijk}$
can be considered as a degenerate
quadric surface (in this case the quadratic equation splits into two
linear factors).
Application of Lemma \ref{lem:8points} to three (degenerate) quadric surfaces 
$\VV_{ij}(\bx)\cup \VV_{ij}(T_k\bx)$, $\VV_{ik}(\bx)\cup \VV_{ik}(T_j\bx)$,
$\VV_{jk}(\bx)\cup \VV_{jk}(T_i\bx)$ and to the fourth one
$\cQ_{ijk}(\bx)$, concludes the proof.
\end{Proof}
\begin{Cor}
The above result can be obviously generalized to quadrilateral
lattices in spaces obtained by intersection of many quadric 
hypersurfaces. Since the spaces of constant curvature, Grassmann manifolds 
and Segr\'e or Veronese varieties
can be realized in this way~\cite{Harris}, the above results can be applied,
in pronciple,  
to construct integrable lattices in such spaces as well. 
\end{Cor}

\section{The Ribaucour transformation}
\label{sec:Ribaucour}
In this Section we suitably adapt the fundamental
transformation of quadrilateral lattices in order to preserve
a given quadratic constraint. Such reductions are called, in the continuous
context, the Ribaucour transformations (see, for example, Chapter~X 
of \cite{Eisenhart-TS}). 

\subsection{The Ribaucour reduction of the fundamental transformation}
\label{sec:sub-Rib}
We first recall (for details, see~\cite{TQL}) the basic results concerning
the fundamental transformation of quadrilateral lattices.
\begin{Th} \label{th:fund}
The fundamental transform $\cF(\bx)$ of the quadrilateral lattice $\bx$
is given by
\begin{equation} \label{eq:fund} 
\cF(\bx) = \bx - \frac{\phi}{\phi_\calC}\bx_\calC \; \; ,
\end{equation}
where\\
i) $\phi:\ZZ^N\ra\RR$ is a new solution of the Laplace equation 
(\ref{eq:Laplace}) 
of the lattice $\bx$\\
ii) $\bx_\calC$ is the Combescure transformation vector, which is a solution 
of the equations 
\begin{equation} \label{eq:DixC} \D_i\bx_\calC = (T_i\sigma_i)\D_i\bx \; \; ,
\end{equation}
where, due to the compatibility of the system (\ref{eq:DixC}) 
the functions $\sigma_i$ satisfy
\begin{equation} \label{eq:Djsi}
\D_j\sigma_i = A_{ij}(T_j\sigma_j - T_j\sigma_i) \; , \quad i\neq j \; ;
\end{equation}
moreover\\
iii) $\phi_\calC$ is a solution, corresponding to $\phi$, of the Laplace 
equation 
of the lattice $\bx_\calC$, i.e.
\begin{equation} 
\label{eq:DiphC} \D_i\phi_\calC = (T_i\sigma_i)\D_i\phi \; \; .
\end{equation}
\end{Th}
\begin{Rem}
Notice that, given $\bx_\calC$ and $\phi$, then equation~(\ref{eq:DiphC})
determines $\phi_\calC$ uniquely, up to a constant of integration.
\end{Rem}
At this point we also recall (see~\cite{TQL} for details) that
an $N$ parameter family of straight lines in $\RR^M$ is called
{\it $N$ dimensional congruence} if any two neighbouring
lines $\gl$ and $T_i\gl$, $i=1,\dots,N$ of the family
are coplanar.The $N$ dimensional quadrilateral lattice $\bx$ 
and $N$ dimensional congruence are called {\it conjugate}, if 
$\bx(\bn)\in \gl(\bn)$, for every $\bn\in\ZZ^N$.
\begin{Cor}
The $N$ parameter family of lines 
$\gl=\langle \bx ,\cF(\bx) \rangle$ forms a congruence, called congruence
of the transformation. Both lattices $\bx$ and $\cF(\bx)$ are 
conjugate to the congruence $\gl$.
\end{Cor}

Theorem~\ref{th:fund} states that in order to construct the fundamental 
transformation of the lattice $\bx$ we need three new ingredients: $\phi$, 
$\bx_\calC$
and $\phi_\calC$. In looking for the Ribaucour reduction of the fundamental 
transformation
we can use the additional information:\\
i) the initial lattice $\bx$ satisfies the quadratic 
constraint~(\ref{eq:quadric}),\\
ii) the final lattice $\cR(\bx)$ should satisfy the same constraint as well.\\
This should allow to reduce the number of the necessary data and, indeed,
to find the Ribaucour transformation it is enough to know the Combescure
transformation vector $\bx_\calC$ only.
\begin{Prop}
The Ribaucour reduction $\cR(\bx)$ of the fundamental transformation of the
quadrilateral lattice $\bx$ subjected to quadratic 
constraint~(\ref{eq:quadric}) 
is determined by the Combescure transformation vector $\bx_\calC$, 
provided that $\bx_\calC$ is not anihilated by the bilinear form $Q$ of the 
constraint
\begin{equation} \label{eq:xCBxC} 
\bx_\calC^t  Q \bx_\calC   \ne 0 \; .
\end{equation}
The functions $\phi$ and $\phi_\calC$ entering in formula (\ref{eq:fund})
are then given by
\begin{align} 
\label{def:phi-q}
\phi & = 2 \bx^t Q \bx_\calC +  \ba^t  \bx_\calC  
\; \; , \\ \label{def:phiC-q}
\phi_\calC & =   \bx_\calC^t Q   \bx_\calC   \; .
\end{align} 
\end{Prop}
\begin{Proof}
Application of the partial difference operator $\D_i$ to the quadratic 
constraint (\ref{eq:quadric}) gives 
\begin{equation} \label{eq:Di-quad}
 (T_i\bx^t) Q  (\D_i\bx)  +  \bx^t Q  (\D_i\bx)
+  \ba^t (\D_i\bx)  = 0 \; \; .
\end{equation}
Applying $\D_j$, $j\ne i$ to equation (\ref{eq:Di-quad}) and making use 
of equations (\ref{eq:Laplace}) and (\ref{eq:Di-quad}), we obtain
\begin{equation} \label{eq:Dij-quad}
 (T_i\D_j\bx^t) Q  (T_j\D_i\bx)  + 
(\D_j\bx^t) Q (\D_i\bx) \;
(1 + T_iA_{ij} + T_jA_{ji}) = 0 \; \; .
\end{equation}
We recall (see \cite{TQL} for details) that, given 
Combescure transformation vector $\bx_\calC$, it satisfies
the Laplace equation
\begin{equation}  \label{eq:LaplaceB}
\D_i\D_j\bx_\calC=(T_{i} A^\calC_{ij})\D_i\bx_\calC+
(T_j A^\calC_{ji})\D_j\bx_\calC \; \; ,\;\; i\not= j, \; \; \;  
i,j=1 ,\dots, N \;  ,
\end{equation} 
with
\begin{equation} \label{eq:Bij}
A^\calC_{ij} = \frac{T_j\sigma_j}{\sigma_i}A_{ij} \; \; .
\end{equation}
We will show that the function $\phi_\calC$, defined in (\ref{def:phiC-q}),
satisfies the Laplace equation (\ref{eq:LaplaceB}) of the lattice $\bx_\calC$. 
Indeed, the difference operator $\D_i$ acting on $\phi_\calC$ gives
\begin{equation*} \label{eq:Dith}
\D_i\phi_\calC =  (T_i\bx_\calC^t) Q   (\D_i\bx_\calC)  + 
 \bx_\calC^t Q (\D_i\bx_\calC)  \; \; .
\end{equation*}
Applying $\D_j$, $j\ne i$ on the above equation and making use of 
the equation (\ref{eq:LaplaceB}) we obtain
\begin{multline} \label{eq:Dij-th}
\D_i\D_j\phi_\calC - (T_iA^\calC_{ij})\D_i\phi_\calC - 
(T_jA^\calC_{ji})\D_j\phi_\calC = \\
=  (T_i\D_j\bx_\calC^t) Q (T_j\D_i\bx_\calC)  + 
 (\D_j\bx_\calC^t)  Q  (\D_i\bx_\calC) \:
(1 + T_iA^\calC_{ij} + T_jA^\calC_{ji}) \; \; .
\end{multline}
Making use of equations (\ref{eq:DixC}) and (\ref{eq:Dij-quad}) we 
transform equation (\ref{eq:Dij-th}) to the form
\begin{multline*} 
\D_i\D_j\phi_\calC - (T_iA^\calC_{ij})\D_i\phi_\calC - 
(T_jA^\calC_{ji})\D_j\phi_\calC = 
(\D_j\bx^t) Q (\D_i\bx)  \times\\
\times \left[ (T_i\sigma_i)(T_j\sigma_j) (1 + T_iA^\calC_{ij} + 
T_jA^\calC_{ji})
- (T_iT_j\sigma_i)(T_iT_j\sigma_j) (1 + T_iA_{ij} + T_jA_{ji}) \right] \; ;
\end{multline*}
the expression in square brackets vanishes due to (\ref{eq:Bij}) and 
(\ref{eq:Djsi}), which shows that the function $\phi_\calC$ does 
satisfy the Laplace equation (\ref{eq:LaplaceB}).

It is easy to see that in order to satisfy constraint (\ref{eq:quadric})
the function $\phi$ must be defined as in (\ref{def:phi-q}). Moreover, 
by direct verification one can check that $\phi$ and $\phi_\calC$
are connected by equation (\ref{eq:DiphC}), which
also implies that $\phi$ satisfies the Laplace equation (\ref{eq:Laplace})
of the lattice $\bx$. 
\end{Proof}
\begin{Rem}
The condition (\ref{eq:xCBxC}) is satisfied, in particular,
when the quadric has non-degenerate and definite bilinear form.
\end{Rem}
Let us discuss the geometric meaning of the algebraic results
obtained above. The congruence $\gl$ of the fundamental transformation
is defined once the Combescure transformation vector is given;
moreover, any generic congruence conjugate to $\bx$ can be obtained 
in this way (for details, see~\cite{TQL}).
The points of the transformed lattice $\cR(\bx)$
belong to the lines of the congruence and to 
the quadric $\cQ$. Therefore, we can formulate the following analogue
of the Ribaucour theorem~\cite{Eisenhart-TS}, 
which also follows directly from Lemma~\ref{lem:8points}.
\begin{Prop}
If a congruence is conjugate to a quadrilateral lattice contained in a 
quadric,
and if each line of the congruence meets the quadric just in two distinct 
points,
then the second intersection of the congruence and the quadric is also a 
quadrilateral lattice conjugate to the congruence.
\end{Prop}
\begin{Rem}
If a line and a quadric hypersurface have non-trivial intersection, then 
they have exactly two points in common 
(counting with multiplicities and points at infinity) or,
alternatively, the line is contained in the quadric.
\end{Rem}

\subsection{Superposition of the Ribaucour transformations}
\label{sec:vect-Rib}
In this Section we consider vectorial Ribaucour transformations,
which are nothing else but superpositions of the Ribaucour
transformations with appropriate transformation data. 

We first recall~\cite{MDS,TQL} the necessary material concerning the
vectorial fundamental transformations.   
Consider $K\geq 1$ fundamental transformations $\cF_k(\bx)$, $k=1,...,K$,
of the quadrilateral lattice $\bx\subset\RR^M$, which are built from
$K$ solutions $\phi^k$, $k=1,...,K$ of the Laplace 
equation of the lattice $\bx$ and $K$
Combescure transformation vectors $\bx_{\calC,k}$, where
\begin{equation*}
\D_i \bx_{\calC,k} = (T_i\sigma_{i,k})\D_i\bx \; , \; i=1,...,N\:, 
\;k=1,...,K\; ,
\end{equation*}
and $\sigma_{i,k}$ satisfy equations
\begin{equation*} \label{eq:Djsik}
\D_j\sigma_{i,k} = A_{ij}(T_j\sigma_{j,k} - T_j\sigma_{i,k}) \; , \quad i\neq 
j \; ;
\end{equation*}
finally, we are given also $K$ functions $\phi^k_{\calC,k}$, which satisfy
\begin{equation*}
\D_i \phi^k_{\calC,k} = (T_i\sigma_{i,k})\D_i\phi^k \; .
\end{equation*}
We arrange functions $\phi^k$ in the $K$ component vector 
$ \bphi = ( \phi^1 ,\dots, \phi^K)^t$,
similarily, we arrange the
Combescure transformation vectors $\bx_{\calC,k}$ into $M\times K$ matrix 
$\bX_\calC = (\bx_{\calC,1}, ... ,\bx_{\calC,K}) $;
moreover we introduce the $K\times K$ matrix 
$\bPhi_\calC= (\bphi_{\calC,1},\dots,  \bphi_{\calC,K})$, whose
columns are the $K$ component vectors 
$\bphi_{\calC,k} = (\phi^1_{\calC,k}, ..., \phi^K_{\calC,k})^t$ 
being the Combescure transforms of $\bphi$
\begin{equation} \label{eq:int-phiC}
\D_i \bphi_{\calC,k} = (T_i\sigma_{i,k})\D_i\bphi \;  .
\end{equation}
\begin{Rem}
The diagonal part of $\bPhi_\calC$ is fixed by the initial
fundamental transformations. To find the off-diagonal part of $\bPhi_\calC$
we integrate equations~(\ref{eq:int-phiC}) introducing $K(K-1)$ arbitrary 
constants.
\end{Rem}
One can show that the vectorial fundamental transformation  ${\boldsymbol 
\cF}(\bx)$
of the quadrilateral lattice $\bx$, which is defined as
\begin{equation} \label{eq:vect-fund}
{\boldsymbol \cF}(\bx) = \bx - \bX_\calC \bPhi_\calC^{-1} \bphi \;  ,
\end{equation}
is again quadrilateral lattice. Moreover, the vectorial transformation
is superposition of the fundamental transformations
\begin{equation*}
{\boldsymbol \cF}(\bx)= (\cF_{k_1}\circ \cF_{k_2}\circ\dots\circ\cF_{k_K} 
)(\bx) \; ,
\quad k_i \ne k_j \quad \text{for} \; i \ne j, 
\end{equation*}
and does not depend on the order in which the transformations are taken. 
In applying the fundamental transformations at the intermediate
stages the transformation data should be suitably transfored as well. 
To prove the 
superposition and permutability statements it is important to notice that
the following basic fact holds:
\begin{Lem} \label{lem:sup-vect-fund}
Assume the following splitting of the data of the vectorial fundamental
transformation
\begin{equation*}
\bphi=\begin{pmatrix}\bphi^{(1)} \\ \bphi^{(2)} \end{pmatrix} \; , \quad
\bX_\calC = \left(\bX_{\calC(1)}, \bX_{\calC(2)} \right) \; , \quad
\bPhi_\calC = \begin{pmatrix} \bPhi^{(1)}_{\calC(1)} &  \bPhi^{(1)}_{\calC(2)} 
\\
\bPhi^{(2)}_{\calC(1)} &  \bPhi^{(2)}_{\calC(2)}  \end{pmatrix} \; ,
\end{equation*}
associated with partition $K=K_1+K_2$. Then the vectorial fundamental
transformation ${\boldsymbol \cF}(\bx)$ is equivalent to the following 
superposition of vectorial fundamental transformations:\\
1. Transformation ${\boldsymbol \cF}_{(1)}(\bx)$ with the data $\bphi^{(1)}$,
$\bX_{\calC(1)}$, $\bPhi^{(1)}_{\calC(1)}$:
\begin{equation*}
{\boldsymbol \cF}_{(1)}(\bx) = \bx - \bX_{\calC(1)} \left( 
\bPhi^{(1)}_{\calC(1)}\right)^{-1}
\bphi^{(1)} \; .
\end{equation*}
2. Application on the result obtained in point 1., 
transformation ${\boldsymbol \cF}_{(2)}$
with the data transformed by the transformation ${\boldsymbol \cF}_{(1)}$ 
as well
\begin{equation*}
{\boldsymbol \cF}_{(2)} ({\boldsymbol \cF}_{(1)}(\bx) ) = 
{\boldsymbol \cF}_{(1)}(\bx) - 
{\boldsymbol \cF}_{(1)} (\bX_{\calC(2)} ) 
\left( {\boldsymbol \cF}_{(1)}   ( \bPhi^{(2)}_{\calC(2)}  )\right)^{-1}
{\boldsymbol \cF}_{(1)} ( \bphi^{(2)} )  \; ,
\end{equation*}
where
\begin{align} 
{\boldsymbol \cF}_{(1)} (\bX_{\calC(2)} ) & = 
\bX_{\calC(2)}  - \bX_{\calC(1)} \left( \bPhi^{(1)}_{\calC(1)}\right)^{-1}
\bPhi^{(1)}_{\calC(2)} \label{eq:F1X2} \\
{\boldsymbol \cF}_{(1)} (\bphi^{(2)} ) & = 
\bphi^{(2)} -  \bPhi^{(2)}_{\calC(1)}  \left( 
\bPhi^{(1)}_{\calC(1)}\right)^{-1}
\bphi^{(1)} \; , \\
{\boldsymbol \cF}_{(1)} ( \bPhi^{(2)}_{\calC(2)}  ) & = 
\bPhi^{(2)}_{\calC(2)} - \bPhi^{(2)}_{\calC(1)} \left( 
\bPhi^{(1)}_{\calC(1)}\right)^{-1}
\bPhi^{(1)}_{\calC(2)} \; \label{eq:F1P22}.
\end{align}
\end{Lem}
\begin{Cor} \label{cor:plan-fund} 
For any $L=0,\dots,K-2$,
the points $\bx^\prime=(\cF_{k_1}\circ \dots\circ \cF_{k_L})(\bx)$, 
$\cF_{k_{L+1}}(\bx^\prime)$, 
$ \cF_{k_{L+2}}(\bx^\prime)$, 
$(\cF_{k_{L+1}}\circ \cF_{k_{L+2}} )(\bx^\prime)$, 
are coplanar. 
\end{Cor}
\begin{Rem} %\label{cor:plan-fund-init} 
The $K(K-1)$ constants of integration in the
off-diagonal part of  $\bPhi_\calC$ are used to construct
``initial quadrilaterals", i.e., the integration constants in 
$\phi^\ell_{\calC,k}$
and $\phi^k_{\calC,\ell}$ ($k\ne \ell$) fix the position of 
$(\cF_{k}\circ \cF_{\ell})(\bx)$ on the plane passing through
$\bx$, $\cF_{k}(\bx)$ and $\cF_{\ell}(\bx)$. The rest of the 
construction is by linear algebra and is the direct consequence
of the geometric integrability scheme (Theorem~\ref{th:constrMQL}).
Any point $\bx$ of the initial lattice, together with its images 
under all possible 
superpositions $\cF_{k_1}(\bx)$,...,$(\cF_{k_1}\circ\cF_{k_2})(\bx)$,..., 
$(\cF_{k_1}\circ\dots\circ\cF_{k_K})(\bx)$, form a network 
of the type of $K$-hypercube. Different paths from $\bx$ to the opposite
diagonal
vertex ${\boldsymbol \cF}(\bx)$ represent various ordering of the fundamental 
transformations
in the final superposition.
\end{Rem}

To find the Ribaucour reduction of the vectorial fundamental
transformation we can use results of Section~\ref{sec:sub-Rib}
to obtain
\begin{align} 
\label{def:phi-qv}
\phi^k & = 2 \bx^t Q  \bx_{\calC,k}  + 
 \ba^t \bx_{\calC,k} 
\; \; , \\ \label{def:phiC-qv}
\phi^k_{\calC,k} & =  \bx_{\calC,k}^t  Q \bx_{\calC,k}  \; .
\end{align} 
Equations (\ref{eq:int-phiC}) and (\ref{def:phi-qv}) lead to
\begin{equation*}
\D_i \left( \phi^k_{\calC,\ell} + \phi^\ell_{\calC,k} \right) = 
(T_i\bx_{\calC,k}^t + \bx_{\calC,k}^t)Q(\D_i\bx_\ell) +
(T_i\bx_{\calC,\ell}^t + \bx_{\calC,\ell}^t)Q(\D_i\bx_k) 
\end{equation*}
which implies that 
\begin{equation} \label{eq:phi-klk}
\phi^k_{\calC,\ell} + \phi^\ell_{\calC,k} =
2 \bx_{\calC,k}^t Q \bx_{\calC,\ell} \; ;
\end{equation}
the constant of integration was found from condition 
$(\cR_k\circ\cR_\ell)(\bx)\subset \cQ$.
\begin{Prop} \label{prop:vect-Rib}
The vectorial Ribaucour transformation ${\boldsymbol \cR}$, i.e., the 
reduction
of the vectorial fundamental transformation (\ref{eq:vect-fund})
compatible with
quadratic constraint (\ref{eq:quadric}), is given by the
following constraints
\begin{align} 
\label{eq:bphi-qv}
\bphi^t & = 2 \bx^t Q  \bX_\calC   + 
 \ba^t \bX_\calC \; \; , \\ 
\label{eq:bPhiC-qv}
\bPhi_{\calC} + \bPhi_{\calC}^t & =  2 \bX_{\calC}^t  Q \bX_{\calC}  \; .
\end{align} 
\end{Prop}
\begin{Proof}
Equations (\ref{eq:bphi-qv}) and (\ref{eq:bPhiC-qv})
are just compact forms of equations (\ref{def:phi-qv})--(\ref{eq:phi-klk}),
which assert that, if $\bx\subset\cQ$ then $\cR_k(\bx)\subset\cQ$, 
$(\cR_k\circ\cR_\ell)(\bx)\subset \cQ$ as well. Moreover, since
Corollary~\ref{cor:plan-fund} still holds, then from Lemma~\ref{lem:8points}
it follows that, at each step of the superposition, the lattice
$(\cR_{k_1}\circ\dots\circ\cR_{k_L})(\bx)$ is also contained
in the quadric, which implies the stated result. 

The algebraic verification that 
${\boldsymbol \cR}(\bx)$ belongs to the quadric $\cQ$
is also immediate. Using condition~(\ref{eq:quadric}) we obtain  
\begin{multline*}
{\boldsymbol \cR}(\bx)^t Q {\boldsymbol \cR}(\bx) + \ba^t \cR(\bx) + c = \\
= \bphi^t(\bPhi_\calC^t)^{-1}\bX_\calC^t  Q \bX_\calC \bPhi_\calC^{-1}\bphi 
- \bx^t Q \bX_\calC \bPhi_\calC^{-1}\bphi -  
\bphi^t(\bPhi_\calC^t)^{-1}\bX_\calC^t Q \bx - \ba^t\bX_\calC 
\bPhi_\calC^{-1}\bphi \; ,
\end{multline*}
which vanishes due to equations (\ref{eq:bphi-qv}) and (\ref{eq:bPhiC-qv}) 
and the following identity 
\begin{equation*}
\ba^t\bX_\calC \bPhi_\calC^{-1}\bphi = \bphi^t(\bPhi_\calC^t)^{-1}\bX_\calC^t 
\ba \; .
\end{equation*}
\end{Proof}
Notice that proving geometrically the above Proposition we proved also
the analog of Lemma~\ref{lem:sup-vect-fund} (which we would like to prove 
algebraically as well).
\begin{Prop} \label{prop:sup-vect-Rib}
Assume the following splitting of the data of the vectorial Ribaucour
transformation
\begin{equation} \label{eq:split-Rib}
\bphi=\begin{pmatrix}\bphi^{(1)} \\ \bphi^{(2)} \end{pmatrix} \; , \quad
\bX_\calC = \left(\bX_{\calC(1)}, \bX_{\calC(2)} \right) \; , \quad
\bPhi_\calC = \begin{pmatrix} \bPhi^{(1)}_{\calC(1)} &  \bPhi^{(1)}_{\calC(2)} 
\\
\bPhi^{(2)}_{\calC(1)} &  \bPhi^{(2)}_{\calC(2)}  \end{pmatrix} \; ,
\end{equation}
associated with partition $K=K_1+K_2$. Then the vectorial Ribaucour
transformation ${\boldsymbol \cR}(\bx)$ is equivalent to the following 
superposition of vectorial Ribaucour transformations:\\
1. Transformation ${\boldsymbol \cR}_{(1)}(\bx)$ with the data $\bphi^{(1)}$,
$\bX_{\calC(1)}$, $\bPhi^{(1)}_{\calC(1)}$.\\
2. Application on the result obtained in point 1. the transformation 
${\boldsymbol \cR}_{(2)}$
with the data 
${\boldsymbol \cR}_{(1)} (\bX_{\calC(2)} )$, 
${\boldsymbol \cR}_{(1)}   ( \bPhi^{(2)}_{\calC(2)}  )$,
${\boldsymbol \cR}_{(1)} ( \bphi^{(2)} )$ given by $\cR$-analogs of 
formulas~(\ref{eq:F1X2})--(\ref{eq:F1P22}).
\end{Prop}
\begin{Proof}
We have to show that the data of both transformations satisfy
constraints~(\ref{eq:bphi-qv}) and 
(\ref{eq:bPhiC-qv}). Since the data~(\ref{eq:split-Rib}) do satisfy 
the constraints we have
\begin{align} 
\bphi^t_{(i)} & = 2 \bx^t Q  \bX_{\calC(i)}   + 
 \ba^t \bX_{\calC(i)} \; \; , \quad i=1,2 \; ,  \\ 
\bPhi_{\calC(i)}^{(i)} + \left(\bPhi_{\calC(i)}^{(i)}\right)^t & =  
2 \bX_{\calC(i)}^t  Q \bX_{\calC(i)}  \; , \\
\bPhi_{\calC(2)}^{(1)} + \left( \bPhi_{\calC(1)}^{(2)}\right)^t & =  
2 \bX_{\calC(1)}^t  Q \bX_{\calC(2)}  \; ,
\end{align} 
this leads immediately to conclusion that the transformation~1. is 
the Ribaucour transfromation. Verification that the data of
the transformation of point~2. satisfy constraints~(\ref{eq:bphi-qv}) and 
(\ref{eq:bPhiC-qv}) can be done by straightforward algebra.
\end{Proof}
\begin{Cor}
Obviously, one can reverse the order of the two transformations (keeping
in mind suitable transformation of their data).
Moreover, the above result implies that assuming a general splitting
$K=K_1+\cdots +K_P$ the final result does not depend on the order
in which the transformations are made.  
\end{Cor}
We finally remark that recursive application of the fundamental 
transformations
can be considered~\cite{TQL} as generating new dimensions of the quadrilateral 
lattice $\bx$. In this context, the Ribaucour transformations generate new
dimensions of the lattice subjected to the quadratic constraint. This 
interpretation
remains valid also in the limit from the quadrilateral lattice $\bx\subset\cQ$
to the multiconjugate net $\bx\subset\cQ$. Therefore the Ribaucour 
transformations
of multiconjugate nets subjected to quadratic constraints generate their
natural, geometricaly distinguished, integrable discrete analogs.

\section{Circular lattices and their Ribaucour transformation}
\label{sec:R-MCL}

In this Section we illustrate the quadratic reduction approach 
on a simple example when the quadric $\cQ$ is
the $M$-dimensional sphere $\SS^M \subset\EE^{M+1}$ 
of radius $1$; the bilinear form
$Q$ is just the standard scalar product ``$\cdot$" in the $(M+1)$ dimensional 
Euclidean space
(we add one dimension for convenience), and the quadratic
constraint~(\ref{eq:quadric}) takes the form $\bx\cdot\bx = 1 $.

\subsection{Circular lattices and the M\"{o}bius geometry}
Given the point $\bN\in\SS^M$ (called the North Pole), consider the 
hyperplane $\TT\simeq\EE^M$ bisecting the sphere and orthogonal to $\bN$. 
In standard way we define the stereographic projection  
$\St:\SS^M  \ra \TT\cup \{\infty \}$ such that for all 
$\bx= ( x^0, \vbx )\in\SS^M\setminus \{ \bN \}$, $\by=\St(\bx)$ 
is the unique intersection point of the line $\langle \bN , \bx \rangle$
with the hyperplane $\TT$:
\begin{align}
\by = \St(\bx) & = \frac{\vbx}{1-x^0} \; , \label{eq:St} \\
\bx = ( x^0, \vbx ) = \St^{-1}(\by)& = 
\left( \frac{|\by|^2 -1 }{|\by|^2 +1} , 
\frac{2\by}{|\by|^2 +1}  \right) \; , 
\quad |\by|^2 =\by\cdot\by \;, \nonumber
\end{align}
and the North Pole is mapped into the infinity point $\infty$.

We recall the basic
property of the stereographic projection~\cite{Pedoe} which is an important 
tool in the conformal (or M\"{o}bius) geometry. 
\begin{Lem} \label{lem:st-circ}
Circles of the sphere $\SS^M$ are mapped in the stereographic projection
into circles or straight lines (i.e., circles passing through the infinity 
point)
of the hyperplane $\TT\simeq\EE^M$.
\end{Lem}
Since the intersection of the plane of any elementary quadrilateral of $\bx$
with the sphere $\SS^M$ is a circle
we have therefore:
\begin{Prop} \label{prop:st-clat}
Quadrilateral lattices in the sphere $\SS^M$ are mapped in the 
stereographic projection into multidimensional circular lattices in $\EE^M$; 
conversely, any circular lattice in $\EE^M$ can be obtained in this way.
\end{Prop}
\begin{Rem} The M\"{o}bius
geometry studies invariants of the transformations of 
Euclidean space, which map circles into 
circles. The M\"{o}bius transformations act therefore within the space
of circular lattices, like the projective transformations act within the space
of quadrilateral lattices (see~\cite{DCN,MQL}). One can identify two circular 
lattices
which are connected by a M\"{o}bius transformation and 
study the circular lattices in the M\"{o}bius geometry approach.
\end{Rem}
Proposition~\ref{prop:st-clat}  provides a
convenient characterization of the circularity constraint~\cite{KoSchief2}.
\begin{Th} \label{th:x2}
The quadrilateral lattice $\by\subset\EE^M$ is circular if and only if the 
scalar function $r=|\by|^2$ is a solution of the 
Laplace equation of the
lattice $\by$.
\end{Th}
\begin{Proof}
The quadrilateral lattice $\by$, satisfying the following system of Laplace 
equations 
\begin{equation}  \label{eq:Laplace-y}
\D_i\D_j\by=(T_i B_{ij})\D_i\by+
(T_j B_{ji})\D_j\by \; \; ,\;\; i\not= j, \; \; \;  i,j=1 ,\dots, N \;  ,
\end{equation} 
is circular if and only if the lattice
$\bx=\St^{-1}(\by)\subset \SS^M \subset \EE^{M+1}$ 
is quadrilateral, i.e., $\bx$ satisfies the Laplace 
equation~(\ref{eq:Laplace}).
Obviously, if $\bx$ is quadrilateral, then the $\EE^M$ part of $\bx$, 
i.e. $\vbx = 2\by/(|\by|^2 +1)$, satisfies
the equation~(\ref{eq:Laplace}) as well. 

The idea of the proof is based on the following observation.
We recall (see~\cite{MQL}) that, if 
$\by$ satisfies 
equations~(\ref{eq:Laplace-y}), then, for any gauge function $\rho$,
the new lattice $\tby=\rho^{-1}\by$ satisfies equations
\begin{equation*}  %\label{eq:Laplace-tx}
\D_i\D_j\tby=(T_i \tB_{ij})\D_i\tby+
(T_j \tB_{ji})\D_j\tby + \tC_{ij}\tbx 
\; \; ,\;\; i\not= j, \; \; \;  i,j=1 ,\dots, N \;  ,
\end{equation*}   
with
\begin{align*}
\tB_{ij}&= (T_j\rho)^{-1}(B_{ij}-\D_j\rho) \; , 
\;\; i\not= j, \; \; \;  i,j=1 ,\dots, N \;  \\
\tC_{ij}&= (T_iT_j\rho)^{-1} ( - \D_i\D_j\rho + (T_i B_{ij})\D_i\rho
+ (T_j B_{ji})\D_j\rho ) \; .
\end{align*}
The rest of the proof follows from the fact that, in our case, 
$\rho = (|\by|^2 +1)/2$ and $\tC_{ij} = 0$. 
\end{Proof}
\begin{Rem}
In the continuous context, the direct analog of Theorem~\ref{th:x2}
leads immediately to orthogonality of the intersecting conjugate
coordinate lines~\cite{Eisenhart-TS}. The above characterization
of circular lattices was postulated in~\cite{KoSchief2} where
its relation to geometry was made via another (equivalent) form of the 
circularity constraint~\cite{DMS}. 
\end{Rem}

\subsection{Ribaucour transformation of the circular lattices}

We recall that the fundamental transformation $\cF(\by)$ of the 
quadrilateral lattice $\by$ generates quadrilateral strip with $N$ 
dimensional basis $\bx$ and transversal direction $\cF$
(the quadrilaterals $\{\by,T_i\by,\cF(\by),T_i\cF(\by)\}$, $i=1,\dots,N$,
are planar as well). When $\by$ is subjected to the circularity condition, 
then it 
is natural to consider only such fundamental transformations which act
within the space of circular lattices~\cite{KoSchief2}.

\begin{Def}
The Ribaucour transformation $\ccR(\by)$ of the circular 
lattice $\by$
is a fundamental transformation such that all the strip with
$N$ dimensional basis $\by$ and transversal direction $\ccR$ is 
made out of circular quadrilaterals.
\end{Def}
\begin{Rem}
It is not enough to define the Ribaucour transformation $\ccR(\by)$ of the 
circular 
lattice $\by$ as a fundamental transformation such that the transformed 
lattice
is circular as well.
\end{Rem}
In this Section we present the Ribaucour transformation
of multidimensional circular lattices from the point of 
view of quadratic reductions. 
Given circular lattice $\by\subset\EE^M$, 
we apply to $\bx=\St^{-1}(\by)\subset\SS^M$ the Ribaucour transformation
$\cR$, defined in Section~\ref{sec:Ribaucour}, obtaining the new lattice
$\cR(\bx)\subset \SS^M$. Since, for points in the sphere, planarity implies 
circularity we conclude that the quadrilaterals $\St(\{\bx,T_i\bx, \cR(\bx),
T_i\cR(\bx)\})$ are circular. This observation, together 
with Lemma~\ref{lem:st-circ} and Proposition~\ref{prop:st-clat}, leads to
the following result.
\begin{Prop} \label{prop:st-Rib}
The transformation $\St(\cR(\St^{-1}(\by)))$ is a Ribaucour transformation
of the circular lattice $\by$; 
conversely, any Ribaucour transformation $\ccR(\by)$ of the circular 
lattice $\by$ can be obtained in this way.
\end{Prop}
\begin{Cor} \label{cor:central-pr}
One can extend, via formula~(\ref{eq:St}), the stereographic 
projection $\St$ to the 
the projection $\opP$ of $\EE^{M+1}$ on $\TT$ with the center 
in $\bN$. In this way the
lines $\gl$ of the congruence of the transformation $\cR$ are
mapped into the lines $\gl^0=\opP(\gl)$ of the congruence of the 
transformation $\ccR$.
However, since the central projection does not preserve parallelism,
it cannot be used directly to define the Combescure
transformation vector $\by_\calC$, from given $\bx_\calC$;
one needs some rescaling.
\end{Cor}
The rest of this Section is devoted to ``algebraization" of the above
geometric observations. 

Consider the circular lattice $\by\subset\EE^M$ and its image in the 
M\"{o}bius
sphere $\bx=\St^{-1}(\by)\subset\SS^M$. The Ribaucour
transformation $\cR(\bx)$ of $\bx$
\begin{equation*}
\cR(\bx) = \bx - \frac{2\bx\cdot\bx_\calC}{ 
\bx_\calC\cdot\bx_\calC}\;\bx_\calC 
\; , %\quad \D_i\bx_\calC = (T_i\sigma_i)\D_i\bx \; ,
\end{equation*}
is mapped in the steroegraphic projection to
\begin{equation*}
\St(\cR(\bx)) = \by - \left( 2\by\cdot\vbx_\calC + 
x^0_\calC\left( |\by|^2 -1 \right) \right)
\frac{x^0_\calC\by + \vbx_\calC}{ |x^0_\calC\by + \vbx_\calC|^2} \; .
\end{equation*}
One can directly verify, that the function
\begin{equation*}
\by_\calC = x^0_\calC\by + \vbx_\calC \; ,
\end{equation*}
is the Combescure transformation vector of the circular lattice $\by$ 
\begin{equation*}
\D_i\by_\calC = (T_i\varrho_i)\D_i\by \; , 
\end{equation*}
with
\begin{equation*}
\varrho_i = x^0_\calC + \frac{2\sigma_i}{|\by|^2+1} \; ,
\end{equation*}
and $|\by_\calC|^2$ satisfies the Laplace equation of the
lattice $\by_\calC$.
Moreover, the function
\begin{equation*}
\psi = 2\by\cdot\vbx_\calC +  x^0_\calC\left( |\by|^2 -1 \right) = 
2\by\cdot\by_\calC -  x^0_\calC\left( |\by|^2 +1 \right) \; ,
\end{equation*}
satisfies equation
\begin{equation} \label{eq:psi-psiC}
\D_i\psi  = \frac{1}{T_i\varrho_i} \D_i(|\by_\calC|^2) = 
\D_i\by \cdot(T_i\by_\calC + \by_\calC) \; . 
\end{equation}
Putting these facts together we arrive to the following characterization
of the Ribaucour transformation of circular lattices~\cite{KoSchief2}
\begin{Th}
The Ribaucour transformation of the circular lattice $\by\subset \EE^M$ 
reads
\begin{equation} \label{eq:Rib-c}
\ccR(\by) = \by - \frac{\psi}{\psi_\calC}\by_\calC \; ,
\end{equation}
where $\by_\calC$ is the Combescure vector of $\by$, 
$\psi_\calC =  |\by_\calC|^2$, and $\psi$ is a solution of 
equation~(\ref{eq:psi-psiC}). 
\end{Th}
We would like to add a few remarks, which follow directly
from the above reasoning, or can be easily verified.
\begin{Cor}
i) When $\opP$ is the projection defined in Corollary~\ref{cor:central-pr}
then
\begin{equation*}
\opP(\bx + \bx_\calC) - \opP(\bx) = \frac{\by_\calC}{1-x^0 - x^0_\calC} \; .
\end{equation*}
ii) The function $\psi$ can be written in the form 
\begin{equation} \label{eq:psi-alg}
\psi = 2  \by \cdot \by_\calC  - (|\by |^2)_\calC \; .
\end{equation}
\end{Cor}
\begin{Cor}
i) In the simplest case, when $\by_\calC=\by$, then $\psi=|\by|^2 - a$,
where $a=\const$, and the corresponding Ribaucour transformation
is the inversion
\begin{equation*}
\ccR(\by) = \cI_a(\by)=a\frac{\by}{|\by|^2} \; .
\end{equation*}
ii) The Combescure transformation of a circular lattice is circular
lattice as well. \\
iii) The Ribaucour transformation can be decomposed into 
superposition of two Combescure 
transformations and inversion: 
\begin{eqnarray*}
\by & \stackrel{\ccR}{\longleftrightarrow} &  \ccR(\by) \\
\calC \updownarrow &   &\updownarrow  \calC \\
\by_\calC  & \stackrel{\cI}{\longleftrightarrow} & 
\cI(\by_\calC) = ( \ccR(\by))_\calC  \; \; .
\end{eqnarray*}
\end{Cor}
\begin{Rem}
We recall that, in the case of the fundamental transformation 
of the quadrilateral lattice $\bx$, the Combescure transformation vectors
$\bx_\calC$ and $(\cF(\bx))_\calC$ (they define the same congruence
but from the point of view of two different lattices)
are related by the radial transformation~\cite{TQL}.
\end{Rem}
For completness, we present also the 
vectorial Ribaucour
transformation of circular lattices. Consider $K$ Ribaucour transformations
of the circular lattice $\by$, which are defined by the Combescure vectors
$\by_{\calC,k}$ and the corresponding transforms $r_{\calC,k}$ of $r=|\by|^2$:
\begin{equation*}
\D_i\begin{pmatrix} \by_{\calC,k} \\ r_{\calC,k} \end{pmatrix}
= (T_i\varrho_{i,k}) \D_i\begin{pmatrix} \by \\ r \end{pmatrix} \; ,
\end{equation*}
which we arrange in $M\times K$ matrix 
$\bY_\calC = (\by_{\calC,1},\dots,\by_{\calC,K})$
and the row vector $\br_\calC = ( r_{\calC,1},\dots , r_{\calC,K} )$.
The corresponding vector $\bpsi$ (see equation~(\ref{eq:psi-alg}))
has components 
\begin{equation*}
\psi^k = 2  \by \cdot \by_{\calC,k}  - r_{\calC,k} \; .
\end{equation*}
Using equations~(\ref{eq:Rib-c}) and (\ref{eq:psi-psiC}) and the condition 
that $(\ccR_k\circ\ccR_\ell)(\by)$
belongs to the circle passing through the points
$\by$, $\ccR_k(\by)$ and 
$\ccR_\ell(\by)$, one can show that
the components of the matrix $\bPsi_\calC$ being defined as
\begin{equation*}
\D_i\psi^{k}_{\calC,\ell} = (T_i\varrho_{i,k})\D_i\psi^k \; ,
\end{equation*}
satisfy condition
\begin{equation*}
\psi^{k}_{\calC,\ell} + \psi^{\ell}_{\calC,k} = 
2 \by_{\calC,k} \cdot \by_{\calC,\ell} \; .
\end{equation*}
Finally, we present the ``circular" analogs of 
Propositions~\ref{prop:vect-Rib} and 
\ref{prop:sup-vect-Rib} of Section~\ref{sec:vect-Rib}.
\begin{Prop} \label{prop:vect-Rib-c}
The vectorial Ribaucour transformation ${\boldsymbol \ccR}$, i.e., 
the reduction
of the vectorial fundamental transformation (\ref{eq:vect-fund})
compatible with the circularity constraint, is given by 
\begin{equation*} \label{eq:vect-Rib-c}
{\boldsymbol \ccR}(\by) = \by - \bY_\calC \bPsi_\calC^{-1} \bpsi \;  ,
\end{equation*}
with the following constraints
\begin{align*} 
\label{eq:bpsi-qv-c}
\bpsi^t & = 2 \by\cdot \bY_\calC   -  \br_\calC \; \; , \\ 
\label{eq:bPsiC-qv-c}
\bPsi_{\calC} + \bPsi_{\calC}^t & =  2 \bY_{\calC} \cdot \bY_{\calC}  \; .
\end{align*} 
\end{Prop}
\begin{Proof}
We have to show that the lattice ${\boldsymbol \ccR}(\by)$ is circular,
i.e., the function $|{\boldsymbol \ccR}(\by)|^2$ is a solution
of the Laplace equation of the lattice ${\boldsymbol \ccR}(\by)$.

First notice that, since $r$ satisfies the Laplace equation of the
lattice $\by$, then the function
\begin{equation*}
{\boldsymbol \ccR}(r) = r - \br_\calC\bPsi_\calC^{-1}\bpsi \;,
\end{equation*}
is a solution
of the Laplace equation of the lattice ${\boldsymbol \ccR}(\by)$.
By straightforward calculations we can verify that 
$|{\boldsymbol \ccR}(\by)|^2={\boldsymbol \ccR}(r)$
\end{Proof}
\begin{Prop} \label{prop:sup-vect-Rib-c}
Assume the following splitting of the data of the vectorial Ribaucour
transformation of the circular lattice $\by$
\begin{equation*} \label{eq:split-Rib-c}
\bpsi=\begin{pmatrix}\bpsi^{(1)} \\ \bpsi^{(2)} \end{pmatrix} \; , \quad
\begin{pmatrix} \bY_\calC \\ \br_\calC \end{pmatrix} = 
\begin{pmatrix} \bY_{\calC(1)} & \bY_{\calC(2)} \\ 
\br_{\calC(1)} & \br_{\calC(2)}  \end{pmatrix} \; , \quad
\bPsi_\calC = \begin{pmatrix} \bPsi^{(1)}_{\calC(1)} &  
\bPsi^{(1)}_{\calC(2)} \\
\bPsi^{(2)}_{\calC(1)} &  \bPsi^{(2)}_{\calC(2)}  \end{pmatrix} \; ,
\end{equation*}
associated with partition $K=K_1+K_2$. Then the vectorial Ribaucour
transformation ${\boldsymbol \ccR}(\by)$ is equivalent to the following 
superposition of vectorial Ribaucour transformations:\\
1. Transformation ${\boldsymbol \ccR}_{(1)}(\by)$ with the data 
$\bY_{\calC(1)}$, $\br_{\calC(1)}$, $\bpsi^{(1)}$, $\bPsi^{(1)}_{\calC(1)}$.\\
2. Application on the result obtained in point 1., 
transformation ${\boldsymbol \ccR}_{(2)}$
with the data
${\boldsymbol \ccR}_{(1)} (\bY_{\calC(2)} )$,
${\boldsymbol \ccR}_{(1)} (\br_{\calC(2)})$, ${\boldsymbol \ccR}_{(1)} 
(\bpsi^{(2)} )$,
${\boldsymbol \ccR}_{(1)} (\bPsi^{(2)}_{\calC(2)}  )$.
\end{Prop}
\begin{Proof}
The reasoning is similar to that of the proof 
of Proposition~\ref{prop:sup-vect-Rib}. The only new ingredient is that
the vector 
\begin{equation*}
{\boldsymbol \ccR}_{(1)} (\br_{\calC(2)}) = \br_{\calC(2)} - \br_{\calC(1)} 
\left( \bPsi_{\calC(1)}^{(1)} \right)^{-1} \bPsi^{(1)}_{(2)} \; ,
\end{equation*} 
consists
of the Combescure transforms of the function
\begin{equation*}
{\boldsymbol \ccR}_{(1)} (r) = r - \br_{\calC(1)} 
\left( \bPsi_{\calC(1)}^{(1)} \right)^{-1} \bpsi^{(1)} = 
|{\boldsymbol \ccR}_{(1)}(\by)|^2 \; .
\end{equation*} 
\end{Proof}

\section{Conclusion and final remarks}

In this paper we presented the theory of quadrilateral lattices
subjected to quadratic reductions. We concentrated our research
on the geometric aspect of the problem of quadratic reductions, 
i.e., our cosiderations concerned the lattice points,
not the corresponding reduction of the MQL equation~(\ref{eq:MQL-A}).
However, it is worth of mentioning that in~\cite{KoSchief2} it was shown that 
the circular lattices, for $N=M=3$, can be described by the discrete BKP
equation~\cite{DJM}. 

The (vectorial) Ribaucour-type transformations of the quadrilateral
lattices in quadrics, also constructed in the paper, 
allow to find new lattices from given ones. In particular, 
a lot of interesting examples can be constructed just applying
the Ribaucour transformations to the trivial background lattices
(see, for example~\cite{MDS,LiuManas}). Moreover, one may expect
that suitable modification of the scheme, based on the
$\bar\partial$-dressing method, applied in~\cite{DMS} to study circular 
lattices, 
can be used to study the quadratic reductions as well.

We conclude the paper with a few general remarks on integrable
lattices.
The multidimensional quadrilateral lattice seems to be quite general 
integrable lattice and other integrable lattices come as their
reductions. Notice~\cite{DCN,MQL} that the quadrilateral lattices
naturally ``live" in the projective space. To obtain reductions
of the quadrilateral lattice one can follow the Cayley and Klein 
approach to subgeometries of the projective geometry, which
was succesfuly applied in~\cite{HolToda,HarmToda} to the (continuous) Toda 
systems. The results of the present paper can be considered
as the basic tool to construct integrable lattices in spaces
obtained by intersection of quadrics. 
As a particular example, we demonstrated here the close connection of
the circular lattices and the M\"obius geometry.

Another way to obtain the integrable reductions of the quadrilateral
lattices (and the corresponding reductions of equation~(\ref{eq:MQL-A})) 
can be achieved by imposing on the lattice special symmetry conditions. 
These additional
requirements may allow for dimensional reduction of 
the geometric integrability
scheme (see examples and discussion in~\cite{RC}). In particular,
the discrete isothermic surfaces~\cite{BP2}, or even the discrete
analogs of the holomorphic functions (see, e.g.~\cite{Schramm} and
references therein) can be considered as further reductions
of the circular lattices.

The third way, pointed out in~\cite{RC}, to obtain new examples of 
integrable lattices
may be to consider quadrilateral lattices (and their reductions)
in spaces over fields different from the field of real numbers.
In particular, geometries over Galois fields (finite geometries)
should give rise to integrable ultradiscrete systems (integrable
cellular automata). 

\section*{Acknowledgements}
The author was partially supported by KBN grant 2P03 B 18509.
The results of Section~\ref{sec:quadratic} were presented on 
the NEEDS Workshop, Colymbari, June 1997.

\end{document}